# The *Rn*-index: a more accurate variant of the *Rk*-index

Alonso Rodríguez-Navarro

*Departamento de Biotecnología-Biología Vegetal, Universidad Politécnica de Madrid, Avenida Puerta de Hierro 2, Madrid 28040, Spain*

*E-mail address:* alonso.rodriguez@upm.es

## 1. Introduction

Rodríguez-Navarro and Brito (2024) recently introduced the *Rk*-index, a non-parametric indicator that reveals the contribution to pushing the boundaries of knowledge—an essential metric for evaluating research in countries and institutions (Rodríguez-Navarro, 2024a). However, the calculation of this contribution is challenging because the involved publications represent only about 0.01% of all papers in a given field (Bornmann et al., 2018; Poege et al., 2019), and are extremely rare even in most advanced countries.

To address this challenge, it was long assumed that common bibliometric indicators, calculated from more frequent publications, could also provide insight into research performance at the breakthrough level (Rodríguez-Navarro & Brito, 2019). However, recent studies have shown that this assumption is incorrect in approximately 50% of cases (Rodríguez-Navarro, 2024a, 2024b). As a result, common bibliometric indicators cannot be reliably used to assess research performance at the frontier of knowledge unless their accuracy at this level has been demonstrated—something that is rarely tested. The *Rk*-index, defined as the geometric mean of the inverses of the global ranks of the 10 most cited papers after adding 20 (Rodríguez-Navarro & Brito, 2024, p. 5; Eq. 1), has been validated using synthetic series.

## 2. Weakness of the *Rk*-index

Despite its general accuracy, the use of the *Rk*-index has revealed a weakness when only a few actors produce most of the advances in a particular research topic. This occurs because, for these actors, the values of the *Rk*-index are close to its maximum value—a situation that was not tested with synthetic series. For example, the *Rk*-index for the domestic papers of the USA in lithium batteries is 32.2, compared to the maximum value of the index, which is 39.5 (Rodríguez-Navarro, 2024c). In such cases, the differences in the *Rk*-index between the main contributors are diminished and fail to reflect the true scientific differences. Another consequence of this weakness is that the summability property of indicators is not fulfilled. Top percentile indicators (Bornmann et al., 2013) satisfy this property. For example, at any top percentile, if institution A has *a* papers and institution B has *b* papers, and if A and B do not share common papers, the combined institution AB will have *a* plus *b* papers. In contrast, any indicator calculated from a limited number of the most cited papers will not fulfill this property in many cases. For instance, in the USA and China (dominant countries) in the fields of solar cells and lithium batteries, the sum of the domestic and collaborative *Rk* and *h* indices is much higher than the *Rk* and *h* indices for the integration of domestic and collaborative papers (Table 1).

Table 1. Summability test for the *Rk* and *h* indices. Values of these indices for the US and China publications in solar cells and lithium batteries, and comparison of the indices for all papers with the sum of the indices for domestic and collaborative papers

| Country | Topic | Domestic | | Collaborative | | All | | Sum of domestic and collaborative | |
|---|---|---|---|---|---|---|---|---|---|
| | | *Rk* | *h* | *Rk* | *h* | *Rk* | *h* | *Rk* | *h* |
| USA | Solar cells | 25.1 | 163 | 19.8 | 185 | 29.2 | 234 | 44.9 | 348 |
| China | Solar cells | 13.1 | 155 | 15.2 | 173 | 20.7 | 207 | 28.3 | 328 |
| USA | Lithium batteries | 32.2 | 169 | 23.2 | 158 | 37.2 | 211 | 55.4 | 327 |
| China | Lithium batteries | 20.6 | 181 | 23.2 | 168 | 27.6 | 220 | 43.8 | 349 |

The values of the *Rk*-indices for domestic and collaborative papers have been taken from Rodríguez-Navarro (2024c)

This weakness of the *Rk*-index prompted the search for similar indicators that are less affected by this limitation.

## 3. The *Rn*-index corrects the weakness of the *Rk*-index

Table 2. Calculation of indicators based on the ranks of most cited papers: *Rk*-index and means of rank ratios. Assessment of China in composite materials

| Rank2[a] | Collaborative publications | | | Domestic publications | | | All publications | | |
|---|---|---|---|---|---|---|---|---|---|
| | Rank1[a] | 1/(20+Rank1) | Rank1/Rank2 | Rank1 | 1/(20+Rank1) | Rank1/Rank2 | Rank1 | 1/(20+Rank1) | Rank1/Rank2 |
| 1 | 1 | 0.048 | 1.000 | 5 | 0.040 | 0.200 | 1 | 0.048 | 1.000 |
| 2 | 3 | 0.043 | 0.667 | 12 | 0.031 | 0.167 | 3 | 0.043 | 0.667 |
| 3 | 4 | 0.042 | 0.750 | 16 | 0.028 | 0.188 | 4 | 0.042 | 0.750 |
| 4 | 9 | 0.034 | 0.444 | 33 | 0.019 | 0.121 | 5 | 0.040 | 0.800 |
| 5 | 17 | 0.027 | 0.294 | 34 | 0.019 | 0.147 | 9 | 0.034 | 0.556 |
| 6 | 19 | 0.026 | 0.316 | 38 | 0.017 | 0.158 | 12 | 0.031 | 0.500 |
| 7 | 35 | 0.018 | 0.200 | 42 | 0.016 | 0.167 | 16 | 0.028 | 0.438 |
| 8 | 36 | 0.018 | 0.222 | 43 | 0.016 | 0.186 | 17 | 0.027 | 0.471 |
| 9 | 45 | 0.015 | 0.200 | 56 | 0.013 | 0.161 | 19 | 0.026 | 0.474 |
| 10 | 47 | 0.015 | 0.213 | 65 | 0.012 | 0.154 | 33 | 0.019 | 0.303 |
| *Rk*-index | 26 | | | 20 | | | 33 | | |
| Geometric mean | | 0.36 | | | 0.16 | | | 0.56 | |
| Arithmetic mean | | 0.43 | | | 0.16 | | | 0.60 | |

[a] Rank1 corresponds to global ranks and Rank2 to local ranks
[b] The most cited papers after combining the publications in the two cases, A and B

The basic approach behind the formulation of the *Rk*-index is the use of the global ranks of the local most cited papers. Based on this approach, different variants of the indicator can be developed, much like the case with the *h*-index (Bornmann et al., 2011).

A simple alternative to directly calculating the *Rk*-index from the global ranks is to calculate the ratios between local and global ranks, and then compute their arithmetic and geometric means to obtain a single comprehensive indicator. Table 2 presents the calculations of the *Rk*-index and the means of the rank ratios in the case of China in the topic of composite materials. In this case the difference between the *Rk*-indices of domestic and collaborative publications is small (26 versus 20) when considering the global ranks of the three most cited papers in the two cases (1, 3, 4 versus 5, 12, 16). Moreover, as expected from the data in Table 1, the deviation from the summability property is substantial; the sum of the individual *Rk*-indices is 40% larger than the *Rk*-index of the combination of domestic and collaborative papers (Table 2). In contrast, the geometric and arithmetic means of the rank ratios correct these weaknesses of the *Rk*-index. These means increase the differences between collaborative and domestic papers from 1.3-fold to 2.3- and 2.7-fold, respectively, and virtually eliminate the deviation from the summability property.

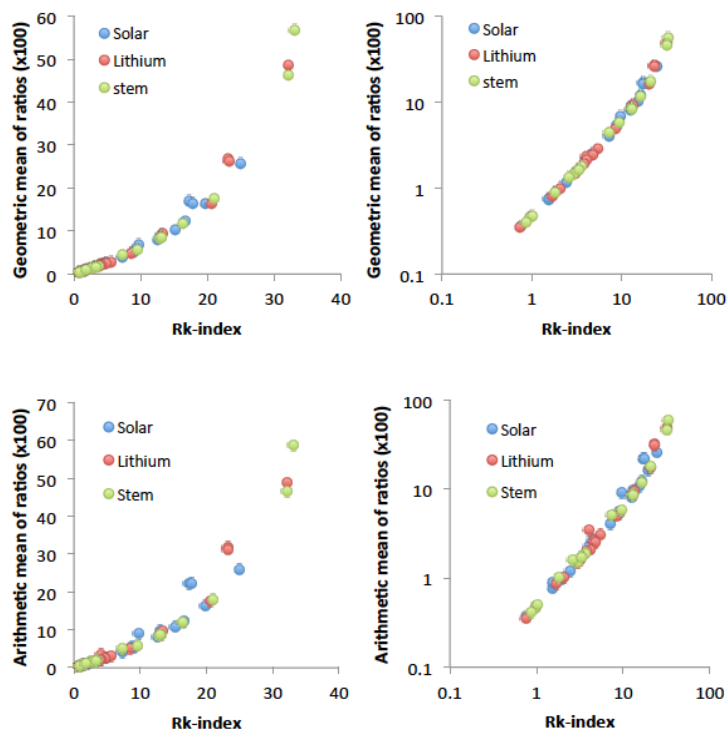

Fig. 1. Relationship between the *Rk*-index and rank ratios: local divided by global rank. Linear and logarithmic plots of the geometric and arithmetic means versus the *Rk*-index.

To further investigate the relationship between the means of the rank ratios and the *Rk*-index, I plotted the corresponding values on both linear and logarithmic scales. Fig. 1 shows these plots for China, the EU, Germany, Japan, Singapore, South Korea, the UK, and the USA, across three topics: lithium batteries, solar cells, and stem cells, considering both domestic and internationally collaborative papers (*Supplementary materials*, Table S1). These plots demonstrate a strong relationship between the *Rk*-index and the two means of the rank ratios. Only six out of 42 data points show a slight but noticeable deviation from the continuous lines that join the points, with these deviations being smaller in the case of the geometric means. These lower deviations were expected because the *Rk*-index is also a geometric mean, suggesting that the deviations arise from differences between the arithmetic and geometric means. Fig. 2 confirms this, as the two means show a perfect linear relationship, except for the aforementioned six data points. This relationship between the means is not surprising, given that the ranks of the most cited papers are not random numbers—they follow a power law or exhibit ordered deviations (Rodríguez-Navarro, 2024b).

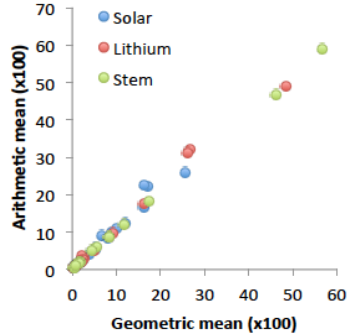

Fig. 2. Relationship between the arithmetic and geometric means recorded in Fig. 1

Based on this finding, the arithmetic mean seemed the most appropriate indicator because it is equivalent to using the sum of the ratios, which is a simple operation. I call this indicator the *Rn*-index (from **R**a**n**k), and it is defined as the sum of the 10 rank ratios multiplied by 10.

The obvious upward curvature of the *Rk*-*Rn*-index plots in the panels with linear scales (Fig. 1), along with the results shown in Table 2, indicates that the *Rn*-index corrects the above-described weakness of the *Rk*-index at high *Rk*-index values. To investigate this further, I calculated the deviations of the *Rn*-index from the summability property in several countries and research topics by comparing the *Rn*-index for all publications with the sum of the *Rn*-indices for domestic and collaborative publications (Table 3). As described earlier (Table 1), deviations from the summability property are to be expected for any indicator based on a limited number of highly cited papers. However, a strong link between these deviations and the magnitude of the indicator, as is seen with the *Rk*-index, seems to be a weakness that depends on the variant of the indicator. Consistent with this reasoning, the data in Table 3 demonstrate that the deviations of the *Rk*-index are random, but also larger for higher values of the index. In contrast, the deviations of the *Rn*-index are only random, without any dependency on high values; for example, in the USA in lithium batteries, the *Rn*-index fulfills the summability property, even though the domestic and collaborative indices are high.

Furthermore, in these tests, the deviations of the *Rn*-index are smaller than those of the *Rk*-index.

Table 3. Summability test for the *Rk* and *Rn* indices. Values of these indices for domestic, collaborative, and all publications in several countries and topics, and deviation of the sum of the indices for domestic and collaborative papers from the corresponding index for all papers

| Country | Topic | *Rk* | | | *Rn* | | | Deviation of sum (%) | |
|---|---|---|---|---|---|---|---|---|---|
| | | Dom[a] | Coll[a] | All | Dom[a] | Coll[a] | All | *Rk* | *Rn* |
| USA | Stem cells | 33.1 | 32.2 | 38.8 | 56.8 | 46.2 | 94.1 | 68.4 | 9.4 |
| USA | Lithium batteries | 32.2 | 23.2 | 37.2 | 48.9 | 31.9 | 82.5 | 48.9 | -2.0 |
| USA | Solar cells | 25.1 | 19.8 | 29.2 | 25.9 | 16.4 | 36.1 | 53.4 | 17.0 |
| China | Lithium batteries | 20.6 | 23.2 | 27.6 | 17.5 | 31.2 | 38.6 | 58.9 | 26.2 |
| China | Solar cells | 13.1 | 15.2 | 20.7 | 9.8 | 10.8 | 17.3 | 37.1 | 18.2 |
| Singapore | Graphene | 7.0 | 7.9 | 11.5 | 4.2 | 4.5 | 7.2 | 30.4 | 20.8 |
| Japan | Lithium batteries | 4.6 | 4.0 | 8.7 | 2.6 | 3.5 | 6.1 | 0.4 | -0.4 |
| South Korea | Semiconductors | 3.8 | 12.2 | 13.5 | 1.9 | 7.9 | 9.0 | 18.1 | 8.6 |
| China | Stem cells | 3.6 | 13.2 | 13.2 | 1.9 | 8.6 | 8.7 | 27.2 | 21.5 |
| Germany | Stem cells | 3.3 | 21.1 | 21.7 | 1.7 | 18.1 | 18.9 | 12.0 | 4.5 |
| Italy | Inflammation | 1.8 | 16.4 | 16.4 | 0.9 | 13.4 | 13.4 | 11.3 | 6.6 |
| India | Solar cells | 1.6 | 1.6 | 2.4 | 0.8 | 0.9 | 1.3 | 27.5 | 28.0 |

[a] Abbreviations: Dom, domestic; Coll, collaborative
[b] The *Rk*-index in domestic and collaborative papers has been taken from Rodríguez-Navarro (2024c)

In summary, the *Rn*-index provides a more accurate measure than the *Rk*-index in countries that produce most of the scientific advancements, thereby correcting the weakness of the *Rk*-index.

**4. Validation of the Rn-index**

Considering the strong numerical relationship between the *Rn* and *Rk* indices (Fig. 1), the validation of the latter with synthetic series (Rodríguez-Navarro & Brito, 2024) extends to the *Rn*-index. However, to provide further support, I studied the correlation between the *Rn*-index and percentile indicators in real cases. The level of important breakthroughs is defined by the top 0.01% of cited papers (Bornmann et al., 2018; Poege et al., 2019), but studying the correlation at this level would require a very large number of global papers in the topic to ensure a sufficient number of papers from each country. This would necessitate the integration of many topics or the use of very wide citation windows. Alternatively, the percentile level can be reduced to the top 0.1%, which seems reasonable, given the results with synthetic series (Rodríguez-Navarro & Brito, 2024). Assuming this, I examined the correlation between the number of top 0.1% cited papers and the *Rn*-index in six research topics: composite materials, graphene, lithium batteries, semiconductors, solar cells, and stem cells. This analysis included all countries in which the number of top 0.1% cited papers was 10 or higher, ensuring statistical robustness (Table 4). In these cases, the two indicators show a clear correlation (Pearson coefficient = 0.65; two-sided p-value = 0.001), though not very high. This can be explained by the fact that the *Rn*-index and the number of top 0.1% cited papers are based on different principles. In the latter all papers included in the percentile have the same value, whereas in the former, each paper has its own value. This loss of information is a weakness of percentile indicators.

Table 4. Correlation between the number of top 0.1% cited papers and the corresponding *Rn*-index in several countries and research topics

| Country (type of papers) | Topic | top 0.1% | *Rn*-index |
|---|---|---|---|
| USA (collaborative) | Stem cells | 39 | 46.6 |
| USA (domestic) | Stem cells | 31 | 58.8 |
| EU (collaborative) | Stem cells | 29 | 32.1 |
| USA (collaborative) | Graphene | 25 | 50.8 |
| China (collaborative) | Graphene | 24 | 42.5 |
| China (domestic) | Graphene | 20 | 21.4 |
| USA (collaborative) | Semiconductors | 17 | 28.1 |
| USA (domestic) | Semiconductors | 16 | 34.5 |
| USA (domestic) | Graphene | 15 | 30.9 |
| USA (domestic) | Solar cells | 15 | 25.9 |
| USA (domestic) | Lithium batteries | 15 | 48.9 |
| USA (domestic) | Composite Materials | 12 | 33.3 |
| China (domestic) | Semiconductors | 12 | 21.4 |
| China (collaborative) | Semiconductors | 12 | 20.6 |
| USA (collaborative) | Solar cells | 12 | 16.4 |
| Germany (collaborative) | Stem cells | 12 | 18.1 |
| USA (collaborative) | Composite Materials | 11 | 32.9 |
| China (collaborative) | Composite Materials | 10 | 43.1 |
| UK (collaborative) | Stem cells | 10 | 11.9 |
| Canada (collaborative) | Stem cells | 10 | 12.4 |
| UK (collaborative) | Stem cells | 10 | 11.9 |

## 5. Conclusions

The contribution to pushing the boundaries of knowledge is a critical metric for evaluating the research performance of countries and institutions, which in many cases is not revealed by common bibliometric indicators (Rodríguez-Navarro, 2024a; 2024b). The *Rk*-index was specifically designed to assess such contributions, and the *Rn*-index is a variant that corrects the weakness of the *Rk*-index, particularly in the evaluation of countries that produce a high proportion of global advancements. This is the case of the USA and China in many technological fields. Additionally, the *Rn*-index is simple to calculate and understand, as it involves only summing the ratios between the local and global ranks of papers, ordered by their citation count. Moreover, the *Rn*-index may also be fractionally counted.

**References**

Bornmann, L., Leydesdorff, L., & Mutz, R. (2013). The use of percentile rank classes in the analysis of bibliometric data: opportunities and limits. *Journal of Informetrics*, *7*, 158-165.

Bornmann, L., Mutz, R., Hug, S. E., & Daniel, H.-D. (2011). A multilevel meta-analysis of studies reporting correlations between the *h* index and 37 different *h* index variants. *Journal of Informetrics*, *5*, 346-359.

Bornmann, L., Ye, A., & Ye, F. (2018). Identifying landmark publications in the long run using field-normalized citation data. *Journal of Documentation*, *74*, 278-288.

Poege, F., Harhoff, D., Gaessler, F., & Baruffaldi, S. (2019). Science quality and the value of inventions. *Science Advances*, *5*, eaay7323.

Rodríguez-Navarro, A. (2024a). Citation distributions and reserach evaluations: The impossibility of formulating a universal indicator. *Journal of Data and Information Science*, *9*, 1-25.

Rodríguez-Navarro, A. (2024b). Uncertain reserach country rankings. Should we continue producing uncertain rankings? *Preprint at arXiv:2312.17560v2*.

Rodríguez-Navarro, A. (2024c). Countries pushing the boundaries of knowlege: the USA's dominance, China's rise, and the EU's stagnation. *Preprint at arXiv:2402.15263*.

Rodríguez-Navarro, A., & Brito, R. (2019). Probability and expected frequency of breakthroughs – basis and use of a robust method of research assessment. *Scientometrics*, *119*, 213-235.

Rodríguez-Navarro, A., & Brito, R. (2024). Rank analysis of most cited publications, a new approach for research assessments. *Journal of Informetrics*, *18*, 101503.